\documentstyle[prd,aps]{revtex}

\input epsf

\begin{document}

\preprint{DUKE-TH-96-104}

\draft

\title{Symmetry Restoration in Gauge Boson Wave Packet Collisions
in Yang-Mills-Higgs Theory}

\author{C. R. Hu, S. G. Matinyan,\thanks{Also at Yerevan Physics 
Institute, Yerevan, Armenia}  B. M\"uller}

\address{Department of Physics, Duke University \\
Durham, North Carolina 27708---0305}

\maketitle

\begin{abstract}
We investigate the restoration of spontaneously broken gauge symmetry 
in collisions of gauge boson wave packets in the SU(2) Higgs model. 
\end{abstract}


\section{Introduction}

The idea of the restoration of spontaneously broken gauge
symmetries under extreme conditions (high temperature, high fermion
densities, strong external gauge fields) was intensely studied in the 
1970's and is well known \cite{1,2} (for review see \cite{3,4}).  Under 
these conditions spontaneously broken gauge theories undergo various 
phase transitions resulting, in most cases, in the restoration of the 
originally broken symmetries.  The consequences of such 
restoration---masslessness of the gauge bosons and fermions, long range 
character of the interactions, etc.---are very important for the 
properties of the very early Universe.  These properties may also play 
a significant role in the context of neutron stars and in the 
multiparticle production at high-energy interactions.

It has long been realized \cite{1,2} that the Lagrangian of the Standard 
Model with a non-zero vacuum expectation value for the scalar field is 
a covariant generalization of the Ginzburg-Landau equation which 
describes a wide variety of ``ordered'' systems (ferromagnets, superfluids,
superconductors, crystals, etc.) and their phase transitions.  From
this point of view very different systems have common general
properties due to their differently revealed broken symmetries.  Using
the analogy with superconductivity, where a strong external electromagnetic 
current (magnetic field) destroys superconductivity resulting in the 
transition to the normal state, an intense current of weakly interacting 
particles (e.g. of $W$-bosons) can restore the gauge symmetry and cause the 
transition to a new phase with massless gauge quanta \cite{2,5}.  

The present paper is devoted to the study of the restoration of the
spontaneously broken gauge symmetry in the collisions of intense,
energetic $W$-boson wave packets.  The paper is organized as follows.
Section II describes the symmetry restoration at zero temperature in the
SU(2) Higgs model.  In Section III this phenomenon is compared with the 
well-known results of symmetry restoration at high temperature.
Section IV is devoted to the stability problem of the one-loop radiatively 
corrected effective potential with the external gauge field in the light 
of the existence of the heavy t-quark.  In Section V we present lattice 
calculations of gauge boson wave packet collisions in which the 
restoration of the SU(2) gauge symmetry is manifested.  Section VI 
contains short concluding remarks.

\section{Restoration of the Broken SU(2) Symmetry by an Intense 
Gauge Field}

This paper is an outgrowth of previous studies of the interactions
of gauge field wave packets in the SU(2) Yang-Mills \cite{6} and 
Yang-Mills-Higgs \cite{7} theories.  Being in essense classical, the
approach of these publications nonetheless reveal interesting 
phenomena associated with the non-perturbative description of 
multiparticle production in collisions of weakly interacting particles.
However, in the present study of symmetry restoration of the scalar
field effective potential as a result of such wave packet collisions, 
this treatment is more adequate since the colliding of intense, energetic 
$W$-boson wave packets serve here as a space-time averaged external source 
exciting the vacuum of the Yang-Mills-Higgs system.  

Our consideration is based on the spontaneously broken SU(2) Yang-Mills 
theory with one Higgs isodoublet coupled to the gauge field.  This 
model retains the most relevant ingredients of the electroweak theory.  
We begin with some preliminaries.

The action describing the model in $(3+1)$ dimensions is given by
\begin{equation}
S = \int d^3xdt \left[ - \textstyle{{1\over 2}} {\rm tr} (F_{\mu\nu}
F^{\mu\nu}) + \textstyle{{1\over 2}} {\rm tr} \left( ({\cal D}_{\mu}
\Phi)^{\dagger} {\cal D}^{\mu}\Phi\right) - \lambda \left( 
\textstyle{{1\over 2}} {\rm tr} (\Phi^{\dagger}\Phi)-v^2\right)^2 \right]~, 
\label{1}
\end{equation}
with ${\cal D}_{\mu}=\partial_{\mu} - ig A_{\mu}^a {\tau^a/2}$,
$\quad F_{\mu\nu} \equiv F_{\mu\nu}^a {\tau^a/2} = 
(i/g) [{\cal D}_{\mu},{\cal D}_{\nu}]$,
and 
\begin{equation}
\Phi = \phi^0 - i\tau^a \phi^a~, \label{3}
\end{equation}
where $\tau^a\;(a=1,2,3)$ are Pauli matrices.  We work in the unitary 
gauge where only physical excitations appear:
\begin{eqnarray}
\Phi &= &(v+\rho/\sqrt{2}) U(\theta)~, \nonumber \\
A_{\mu} &= &U(\theta) W_{\mu} U^{-1}(\theta) - {1\over ig}\left(
\partial_{\mu} U(\theta)\right) U^{-1}(\theta)~, \label{4}
\end{eqnarray}
with $U(\theta) = \exp (i\tau^a\theta^a)$ being a local gauge rotation.

The real field $\rho$ describes the oscillations of the scalar field about 
its vacuum expectation value $v$, and $W_{\mu}$ is the field of the gauge 
boson.  $W_{\mu}$ and $\rho$ obey the following classical equations of motion:
\begin{eqnarray}
[{\cal D}_{\mu},F^{\mu\nu}] + M_W^2W^{\nu} + {1\over\sqrt{2}} g^2 v\rho
W^{\nu} + {1\over 4}g^2\rho^2 W^{\nu} &= &0, \label{5} \\
\left( \partial_{\mu}\partial^{\mu}+M_H^2\right)\rho + 3\sqrt{2}
\lambda v\rho^2 + \lambda\rho^3 - {\sqrt{2}\over 4} g^2v W_{\mu}^a
W^{a\mu} - {1\over 4}g^2\rho W_{\mu}^aW^{a\mu} &= &0~. \label{6}
\end{eqnarray}
${\cal D}_{\mu}$ and $F^{\mu\nu}$ are defined in terms of $W^{\mu}$.
$M_H= 2v\sqrt{\lambda}$ and $M_W = gv/\sqrt{2}$ are the masses of Higgs
and gauge bosons, respectively.\footnote{Note that the gauge field acts 
as a source for Higgs field excitations in (\ref{6}).  This permits us 
to consider the $W$-field classically with respect to the excitation of the
scalar field.  The equation (\ref{5}) for gauge field does not possess a 
source term.} \footnote{Our normalization of the $\rho$-oscillations about 
$v$ in (\ref{4}) corresponds to the choice $v$ = 174 GeV.  Thus, $G_F/\sqrt{2} 
= 1/4v^2$, where $G_F$ is the effective four-fermion coupling constant.}

It is easy to see that after a scaling transformation
\begin{equation}
x'_{\mu} = gvx_{\mu},\quad \Phi' = \Phi/v, \quad A'_{\mu} = A_{\mu}/v~,
\label{7}
\end{equation}
the action (\ref{1}) and the equations of motion (\ref{5}) and (\ref{6}) 
possess only a single parameter $\lambda'\equiv \lambda/g^2 = 
M_H^2/8M_W^2$.  However, since in the following simulation of the 
colliding wave packets there are other parameters involved in the 
initial conditions, we consider the unscaled equations (\ref{5}) and
(\ref{6}) in more detail.

Equation (\ref{6}) describes the excitations of the scalar field about
the Higgs vacuum $|\Phi| = v$.  It has a second exact solution $\rho = 
-\sqrt{2}v$ (corresponding to $\vert\Phi\vert =0$) with $W^a_{\mu}$ being
arbitrary.  For this solution the last three terms in (\ref{5}) cancel 
each other and eq. (\ref{5}) describes massless gauge bosons.

In terms of excitations $\chi=\rho+ v\sqrt{2}$ around $\vert\Phi\vert=0$,
the equations (\ref{5}) and (\ref{6}) assume the form:
\begin{eqnarray}
[{\cal D}_{\mu},F^{\mu\nu}] + {1\over 4} g^2 \chi^2 W^{\nu} &= &0~, \label{8} \\
\left[\partial_{\mu}\partial^{\mu}-\frac{M_H^2}{2}  \left( 1 + {g^2W^2 
\over 8\lambda v^2}\right)\right] \chi +\lambda \chi^3 &= &0~. \label{9}
\end{eqnarray}
The expression $g^2W^2/8\lambda v^2$ in (\ref{9}), where $W^2(x) \equiv 
W_{\mu}^a W^{a\mu}$, plays an important role:  Depending on the sign and 
magnitude of $W^2(x)$, the new ``ground'' state $\vert\Phi\vert = 0$ 
can be unstable or stable.  For positive $W^2(x)$  this state is obviously 
unstable noting that the effective mass of the scalar excitations described 
by (\ref{9}) is imaginary.

For time-like $(k^2>0)$ gauge bosons $W^2(x)$ is always negative.  In 
particular, this is so for transverse gauge bosons for which $W_0^a=0$. 
For space-like $(k^2 <0)$ gauge bosons $W^2(x)$ can be of both signs.  On 
the other hand, the luminosity of transversely polarized gauge bosons in 
proton-proton collisions at high energy, under realistic conditions, is 
much higher than that of longitudinally polarized ones and increases with 
energy \cite{8}.  Of course, the general suppression by a factor $M_W/E$ 
of the  emission amplitude of longitudinal $W$-bosons from a quasi-massless 
fermion is compensated by the nonvanishing of this amplitude in the
forward direction in contrast to that for transverse $W$-bosons leading,
in general, to the dominance of the cross-sections for longitudinal
$W$-$W$ interactions over transverse ones \cite{8}.  But, as seen from the 
above considerations, for the restoration of symmetry the important 
factor is not the cross section of $W$-$W$ interaction, but rather the 
intensity of the ``pulse'' $W^2$ which appears in equation (\ref{9}) as 
an external source (cf. the footnote 1).

Concerning the $W$-$W$ collisions, in the following we will consider 
collisions of transverse gauge bosons for which the relation $\partial^{\mu} 
W_{\mu}^a = 0$ holds.  This also allows us to work in the temporal gauge 
$W_0^a=0$, which is convenient for lattice calculations in the Hamiltonian 
formulation (see Section V for details).  We would like to emphasize that, 
considering the case of transversely polarized $W$-$W$ scattering, we exclude 
the important issue of the potentially strong longitudinal $W$-$W$ interaction 
which may exist beyond the Standard Model \cite{9}.  The restoration of the 
broken symmetry in the transverse $W$-$W$ collisions studied here is a 
phenomenon entirely within the confines of the Standard Model.

Since the colliding wave packets are highly energetic $(\bar k\gg M_W,
M_H)$ and rapidly oscillating in space and time, the Higgs field
oscillations will only follow effectively the average of the gauge wave 
packets.  We therefore introduce the space-time averaged quantity
\begin{equation}
\eta \equiv {{g^2\langle (W^a_i)^2 \rangle}\over{8\lambda v^2}} = 
{{M_W^2}\over{M_H^2}}\; {{\langle (W^a_i)^2 \rangle}\over{v^2}}~. \label{10}
\end{equation}
Equation (\ref{9}) describes the excitations of the scalar field $\chi$ 
with an effective mass 
\begin{equation}
\tilde M_H^2 = {M_H^2\over 2}(\eta-1)~. \label{11}
\end{equation}
For $\eta<1$ the excitations around the $\Phi=0$ ground state is
tachyonic, indicating instability.  For $\eta >1$, however, these 
oscillations are stable.  In particular, for the spatially homogeneous 
case there exists an exact stable solution of (\ref{9})
\begin{equation}
\chi(t) = {\chi_0 e^{i\tilde M_H t} \over 1 - {\lambda\over 8\tilde
M_H^2} \chi_0^2 e^{2i\tilde M_H t}}~, \label{12}
\end{equation}
where $\chi_0$ is a constant \cite{10}.

It is instructive to analyze the effective potential $V(\chi,\eta)$
for the excitations $\chi$ described in equation (\ref{9}):
\begin{equation}
V(\chi,\eta) = -\lambda v^2 (1-\eta) \chi^2 + {\lambda\over 4} \chi^4~. 
\label{13}
\end{equation}
Equation (\ref{13}) has two different {\it stable} minima, depending
on the magnitude of $\eta$:
\begin{eqnarray}
\hbox{For $\eta < 1$,} \quad &&\chi_{\rm min} = \pm \sqrt{2}v(1-\eta)^{1/2}, 
\quad {\rm i.e.}\; \vert\Phi\vert = v(1-\eta)^{1/2}~. \label{14} \\
\hbox{For $\eta \ge 1$,} \quad &&\chi_{\rm min} = 0, \quad {\rm i.e.}\; 
\Phi =0~.  \label{15}
\end{eqnarray}

Stable excitations about these ``vacua'' have the following squared
masses:
\begin{eqnarray}
\tilde M_H^2 &= &{M_H^2\over 2} \vert 1-\eta\vert [1+\theta(1-\eta)]
\label{16}~, \\
\tilde M_W^2 &= &M_W^2 (1-\eta) \theta (1-\eta)~. \label{17}
\end{eqnarray}
Thus for $\eta>1$, the broken SU(2) gauge symmetry is restored and
oscillations of the scalar field occur about the symmetrical state
$\vert\Phi\vert=0$, not about $\vert\Phi\vert=v$.  The effective mass
of the gauge bosons in the region where the symmetry is restored
$(\eta>1)$ vanishes.  For $\eta <1$, the vacuum is changed gradually as
$(1-\eta)^{{1\over2}}$, approaching $\vert\Phi\vert=0$.  For $\eta<1$, 
the ratio $r=\tilde M_H/\tilde M_W = M_H/M_W$ has no dependence on $\eta$.

\section{Comparison With The High Temperature Case}

The above described features are characteristics of a second order
phase transition.  From this point of view, it is illuminating to 
compare the restoration of symmetry in an intense gauge field associated
with time-like $W$-bosons $(k^2>0$, or $W_{\mu}^aW^{a\mu}<0)$ with the 
analogous phenomenon occurring at high temperature.

For the same SU(2) Higgs model the parameter governing the restoration
of symmetry at finite temperature at one-loop level is \cite{11,12,13}
\begin{equation}
\eta_T = {T^2\over v^2} \left( {1\over 3} + {M_W^2\over M_H^2}\right)~.
\label{18}
\end{equation}

The high temperature acts here in the same way as an intense time-like 
gauge field (or as an extra-high fermion current density \cite{3,4}).  
The joint action of these two effects, of course, would make the 
transition easier.  It is important to stress that our essentially 
classical approach is not reliable for $\eta$ close to unity where the
radiative corrections are significant even at small coupling constants.
However, it is expected that the radiative corrections are not important 
in the region described by the inequality \cite{3}.
\begin{equation}
\vert 1-\eta \vert \ge \lambda,~g^2~. \label{19}
\end{equation}

\section{On the Role of the Radiative Corrections at one Loop Level}

Evidence of the existence of the very heavy $t$-quark $(m_t \approx$
175 GeV) \cite{14} raises the question of the stability of the ground 
state with and without external influence.  In this section
we show that the possible inclusion of heavy fermions into our
radiatively corrected effectived potential (\ref{13}), under realistic
conditions outlined in Section 3, does not change the resulting
picture for moderate $\eta$ and the perturbative range of $\lambda$ and
$g^2$.

The radiatively corrected potential (\ref{13}) has a well-known contribution 
from one loop diagrams including the physical SU(2) gauge bosons, Higgs 
scalar, and $t$-quark \cite{15} (we ignore other much lighter quarks and 
other scalars).  We have for our choice of the normalization of the 
$\chi$-excitations and of $v$:
\begin{equation}
V_c(\chi,\eta) = V(\chi,\eta) + {C\over 4} \chi^4 \ln {\chi^2\over 2M^2}~, 
\label{20}
\end{equation}
where $M$ is an arbitrary normalization scale which will be chosen in
the following equal to $v$.  For $C$ we have \footnote{The unconventional
numerical factor in $C$ is due to our normalization of the
$\rho$-oscillations in (\ref{4}) and our choice of $v=$ 174 GeV.
Summation in (\ref{21}) runs over gauge bosons $W^{\pm}$ and $Z$.}:
\begin{equation}
C = {1\over 256 \pi^2v^4} \left[ \sum_V 3M_V^4 + M_H^4 - 4M_t^4
\right]~. \label{21}
\end{equation} 

Inserting into (\ref{21}) the physical masses of particles, we are
left with
\begin{equation}
C = 4\times 10^{-4} \left[ \left( {M_H\over v}\right)^4 - 3.5\right]~.
\label{22}
\end{equation}
As we are interested in the possible destablizing role of the
$t$-quark, we consider the case $C<0$, i.e. $M_H <$ 238 GeV, or,
$\lambda < 0.47$.  On the other hand, $\lambda > 0.03$ follows from
the present experimental lower bound on the Higgs mass $(M_H >65.2$ GeV) 
\cite{16}.

Now, the minimum of the full effective potential (\ref{20}) is
\begin{equation}
\chi_{\rm min}^2 = {2v^2 (1-\eta) \over 1 + {C\over\lambda} \left[
\ln {v^2(1-\eta)\over M^2} + {1\over 2}\right]}\cdot \theta (1-\eta)~. \label{23}
\end{equation}
The term ${1\over 2}$ in the denominator of (\ref{23}), of course, can be
absorbed into $M$.  The prefactor of the logarithm in (\ref{23}) is small
({${{\vert C\vert}\over\lambda} <0.05$).

The condition for the validity of the one-loop approximation (we take 
$M\sim v$) is:
\begin{eqnarray} 
1-\eta &> &e^{-{\lambda\over\vert C\vert}},\quad {\rm for}~ \eta< 1 \\
\eta -1 &< &e^{\lambda/\vert C\vert}, \quad {\rm for}~ \eta >1 \label{24}
\end{eqnarray}
which is well satisfied considering our restriction (\ref{19}) of the
whole reliability of the approach and for not very large $\eta$.  Thus we 
conclude that the picture does not change drastically by the inclusion of 
the radiative corrections with the dominated $t$-quark contribution.

Indeed, one has for $\vert\eta - 1\vert > \lambda,g^2$ and not large
$\eta$:
\begin{eqnarray}
{\vert\Phi\vert}_{\rm min} &= &v(1-\eta)^{1/2} \left[ 1 - {C\over 2\lambda} \ln
(1-\eta)\right] \theta(1-\eta)~, \nonumber \\
\tilde M_H^2 &= &{M_H^2\over 2} \vert 1-\eta\vert \left( 
1-{C\over\lambda}\right) \left( 1+\theta (1-\eta)\right)~, \nonumber \\
\tilde M_W^2 &= &M_W^2 (1-\eta) \left( 1 - {C\over\lambda}\ln
(1-\eta)\right) \theta(1-\eta)~,  \nonumber \\
\lambda_c &\equiv & {1\over 6} {d^4V\over d\chi^4}
\bigg\vert_{\chi=\chi_{\rm min}} =
\lambda \left[ 1 + {C\over\lambda} \left(\ln \vert 1-\eta\vert + 
{23\over 6}\right) \right]~. \label{25}
\end{eqnarray}

From (\ref{25}) it is seen that the use of the bare coupling constant
$\lambda$ in all previous estimates is reasonable.

The physical reason for this small influence of the $t$-quark is
clear: the choice of moderate external vector field amplitude 
$W^2(<0)$ ensures the reliability of one-loop approximation and, as we 
see, the radiative corrections cannot stop the tendency towards 
the restoration of the symmetry.

It is worthy to emphasize that whereas our arguments are well grounded 
for the $\eta< 1$ region, relations (\ref{25}) have only qualitiative 
validity even for moderate $\eta$ in the region $\eta >1$ where infrared 
singularities connected with the masslessness of the gauge bosons are 
important and practically incurable.  These difficulties are common for 
phase transitions in many-body problems treated by perturbation theory or 
by mean field methods \cite{3}.

Note, finally, that in the potential (\ref{20}) the renormalization
scale dependence of its parameters is ignored.  However, the one-loop 
potential (\ref{20}) has all the qualitative features of the ``full'' 
renormalization group improved potential, and the results thus obtained 
will not be substantially changed if we use the ``exact'' potential \cite{17}.

\section{Restoration of the SU(2) Gauge Symmetry in Gauge Wave Packet 
Collisions}

As follows from the arguments in Section IV, the reliability of our 
analysis is confined to the values of not very large $\eta$.  For $\eta\gg 
1$ our treatment is, in principle, not well grounded, but that occurs only
in the very early Universe.  On the other hand, the study of the gradual 
decrease of the vacuum expectation value as a function of $\eta$ from 
$\vert\Phi\vert = v$ to $\vert\Phi\vert = v(1-\eta)^{1/2}$ for $\eta <
1$ is free from the difficulties mentioned in Section IV.

In this section we study the collision of gauge wave packets from the 
point of view of the restoration of the SU(2) broken symmetry.  In \cite{7} 
we have investigated the evolution of the gauge and Higgs's field as a 
function of the colliding wave packet amplitudes and of the mass ratio 
$r\equiv M_H/W_W$.  For a wide range of the essential parameters we have 
observed collisions of the initial wave packets resulting in final 
configurations with dramatically different momentum distributions.  
Note that the classical approach used in these calculations, although 
suffering from the common ultraviolet catastrophe of classical field 
theory revealed in the momentum distribution of the final states (see 
\cite{7}), is adequate in the present context since we are not interested 
in the final states.

Our numerical study here, as in \cite{7}, is based on the Hamiltonian
formulation of lattice gauge theory \cite{18} (see \cite{19,6} for
more details), in which the dynamical variables are link variables
defined as
\begin{equation}
U_{\ell} = \exp \left( -iga A_{\ell}^c \tau^c/2\right)~. \label{26}
\end{equation}
($\ell$ is the link index.)  As in \cite{6,7} we work with a one-dimensional
lattice of size $L=Na$, where $N$ is the number of lattice sites and $a$ 
the lattice spacing.  As mentioned in Section II, we ``collide'' 
transverse $W$-bosons to ensure the negativity of $W_{\mu}^aW^{a\mu}$ as 
well as the temporal gauge condition $W_0^a=0$.  

The initial configuration is given by two well-separated left- and 
right-moving wave packets originally centered at $z_{L(R)}$ with average 
momenta $k=(0,0,{\bf k})$ and width $\Delta k$.  $z_{L(R)}$ are 
chosen so that the wave packets are positioned symmetrically about the 
center of the lattice:
\begin{eqnarray}
W^{c,\mu} &= &W_{\rm L}^{c,\mu} + W_{\rm R}^{c,\mu}~, \nonumber \\ 
W_{\rm L}^{c,\mu} &= &(0,0,1,0)n_{\rm L}^c \psi(z-z_{\rm L},-t)~,
\nonumber \\
W_{\rm R}^{c,\mu} &= &(0,0,1,0) n_{\rm R}^c \psi(z-z_{\rm R}, t)~,
\label{27}
\end{eqnarray}
with $n_{\rm L}^c$ and $n_{\rm R}^c$ being the polarization vectors in
isospin space.

A right-moving wave packet centered at $z=0$ at $t = 0$ with mean wave
number $\bar k$, mean frequency $\bar\omega = (\bar k^2+M_W^2)^{1/2}$
and width $\Delta k$ is described by
\begin{equation}
\psi(z,t) = {\hbar^{1/2}\over \sqrt{\pi^{3/2}\Omega\Delta k\sigma}}
\int_{-\infty}^{+\infty} dk_z\; e^{-(k_z-\bar k)^2/2 (\Delta k)^2} 
\left[ \cos (\omega t - k_zz)\right] 
\label{28}
\end{equation}
with $\omega = \sqrt{k_z^2 +M_W^2}$, and
\begin{equation}
\Omega = \bar\omega \left[ 1+ {1\over 4} \left( 1- \Big( {\bar k
\over\bar\omega}\Big)^2 \right) \Big({\Delta k \over\bar\omega}
\Big)^2 + O \Big( {\Delta k \over\bar\omega}\Big)^4\right]~, 
\label{29}
\end{equation}
where the amplitude is fixed by requiring energy equal to
$\hbar\bar\omega$ per cross-sectional area $\sigma$.  (In the following
we set $\hbar = 1$.)  From (\ref{28}) we obtain
\begin{equation}
\psi\vert_{t=0} = \sqrt{{2\Delta k \over\sqrt{\pi}\Omega\sigma}}\;
e^{-(\Delta kz)^2/2} \cos(\bar k z)~, \label{30}
\end{equation}
and 
\begin{equation}
\frac{d}{dt}{\psi} |_{t=0}=\sqrt{2\bar \omega\Delta k\over\sqrt{\pi}\sigma (\Omega/\bar\omega)} 
e^{-(\Delta k z)^2/2} \left[\sin(\bar kz) +{\bar k\over\bar\omega}
{\Delta k\over\bar\omega}\Delta k z \cos(\bar kz)+{\cal O}\left( \Big({\Delta k\over\bar \omega}\Big)^2 \right)
\right]~.    \label{30a}
\end{equation}
Initial condition for the Higgs field is the vacuum solution:
\begin{equation}
\phi^0 = v, \quad \phi^a=0, \quad \dot\phi^0=\dot\phi^a =0 \quad {\rm
at} \quad t=0~. \label{31}
\end{equation}

To determine the number of independent parameters, it is useful to
rescale (\ref{30}) according to (\ref{7}) in the high-energy limit
$(\Omega \approx \bar\omega\approx\bar k)$:
\begin{equation}
\psi'\vert_{t'=0} = {1\over v} \psi\vert_{t=0} \approx \sqrt{{2\Delta k 
\over \sqrt{\pi}\bar k\sigma v^2}}\; e^{-\left({\Delta k \over gv} z'
\right)^2\Big/ 2} \cos\left( {\bar k \over gv} z'\right)~,
\label{32}
\end{equation}
which contains three dimensionless parameters:  $\bar k/M_W$, $\Delta
{\bar k}/M_W$, and $\sigma M_W^2/g$ \cite{7}.  There appears one 
more parameter in the initial conditions---the relative initial isospin
orientation of the colliding wave packets, $\theta_c$.  As shown in 
\cite{6,7}, the parameter $\theta_c$ plays an important role in the 
collisions:  Whereas for $\theta_c \not=0$ (essentially non-abelian 
configuration) collisions of the wave packets resulted in highly inelastic 
final states, for parallel isospins (pure abelian configuration) the 
collisions were practically elastic.  However, for the phenomenon we are 
interested in here---restoration of the broken symmetry---the coupling 
between the gauge and Higgs field is essential and not the relative 
orientation of the gauge bosons in isospin space. 

It is convenient to have a scale parameter in the dimensionless 
parameters $v$ rather than $M_W$.  Combining equations of motion and 
initial conditions, we finally have four independent parameters:
\begin{equation}
r\equiv {M_H\over M_W}, \; {\bar k\over v}, \;{\Delta k \over v}, \;
{\rm and} \; {\sigma v^2\over g^2}~. \label{33}
\end{equation}
In our simulations we always chose $\bar k \gg M_W$ and $\bar k \gg
\Delta k$ to model high energy collision and to have wave packets as
well-defined objects.  The gauge coupling constant was fixed to $g= 0.65$.

We estimated the parameter $\eta$ using two different configurations
of the gauge boson wave packets.  Firstly, we consider the configuration 
of the two incident wave packets at the space-time point when they 
overlap completely.  Then from (\ref{10}),(\ref{27}), and (\ref{28}) we have
\begin{equation}
\eta = \cases{{4\over r^2} \kappa^2~, &for parallel isospin orientation \cr
{2\over r^2} \kappa^2~, &for orthogonal isospin orientation \cr
0~, &for antiparallel orientation \cr} \label{34}
\end{equation}
where $\kappa = {2\Delta k\over \sqrt{\pi}\bar k\sigma v^2}$ is just the
parameter we have introduced earlier (see (\ref{31}) in \cite{6}).  As 
found there, for $\kappa\ge 1$ the $W$-$W$ wave packet collisions produced 
strongly inelastic final states (for non-parallel isospin orientation).
However, one should note that according to Eq. (\ref{6}) the scalar 
field reacts to the gauge field long before the overlapping of the wave 
packets (see also figures below). That makes the behavior of the collisions
less sensitive to the relative orientations of the wave packets in isospin
space. Practically, as seen in our simulations, there is no difference 
between the parallel and the antiparallel cases.

Let us estimate $\eta$ taking the initial $(t=0)$ configurations of two 
separate wave packets.  From (\ref{10}),(\ref{28}), and (\ref{33}) we 
have after spatial averaging
\begin{equation}
\eta = {\kappa^{\prime 2}\over r^2} \equiv {2\Delta
k\over \bar k\sigma v^2r^2}~. \label{35}
\end{equation}
Note that no isospin factor appears here after spatial averaging.  As
we see, the two estimates (\ref{34}) and (\ref{35}) almost coincide.
A sufficiently large $\eta$ is achievable for realistic values
of $r>1$ and permissible ratio ${\Delta k}/{\bar k}$ if we take
$\sigma < 1/v^2$.

The natural spread of the $W$-boson wave packets emitted by fast moving
fermions (quarks, leptons) is of order $v$ in the comoving reference
frame.  Therefore, the transverse area of the $W$-boson wave packet is
not larger than $1/v^2$.  Of course, the more reliable estimate of
$\eta$ depends on the detailed shape of the wave packet
and requires a full three-dimensional analysis.

Figures 1 and 2 describe the results of our lattice simulations of
the collisions of $W$ wave packets.  In view of the above mentioned relation 
between the parameters $\eta$ and $\kappa$ (or $\kappa'$) (see (\ref{34}) 
and (\ref{35})) we present here only results for wave packets with
orthogonal isospins since parallel isospin orientation leads qualitatively 
to the same symmetry restoration patterns.

In the first column of Figure 1, the space development of the colliding
$W$ wave packets with $\eta = 1.32$ is shown for different times.  As seen 
from this figure, at time $t \approx 300$ the $W$ wave packets collide and 
then begin to separate.  Just about at this time one expects to observe 
the restoration of symmetry, i.e. the oscillations of the scalar field 
$\vert\Phi\vert$ about a new ground state located below the ``old'' vacuum 
$\vert\Phi\vert/v=1$.  After the separation of the wave packets $(t>300)$ 
the excitations of the scalar field tend again to oscillate about the 
``old'' vacuum, i.e. the gauge symmetry is broken again.  This is seen 
clearly in the second and third columns of Figure 1.  The second column 
shows the space-time evolution of the magnitude of the Higgs field 
$\vert\Phi\vert/v$.  The third column shows the Higgs field smoothed over 
50 lattice sites, suppressing the small scale fluctuations, in order to
facilitate a comparison with the definition (\ref{10}) of the effective
parameter $\eta$ in terms of the averaged strength of the $W$-boson field.  
At the 
moment of interaction, $t\approx 300$, the lowering of the average value 
of the Higgs field at the location of the coalescing wave packets is clearly 
visible.

In Figure 2, we show the $\eta$ dependence of the $W$-$W$ collisions (first
column) and the corresponding behavior of the scalar field (second and
third columns) at $t=300$.  To change $\eta$ we vary the amplitude of wave
packets by changing $\sigma$;  all other parameters are kept fixed.
It is evident that no significant symmetry restoration occurs for $\eta <1$, 
as expected.  For $\eta \gg 1$ the situation becomes more complicated, 
probably because already a single wave packet can restore the symmetry in 
this case.  It is seen, however, that the spatial region with restored 
symmetry becomes wider as the parameter $\eta$ grows.

\section{Concluding Remarks}

The above study of colliding gauge wave packets in the spontaneously
broken SU(2) Yang-Mills-Higgs model visualizes the process of the SU(2)
gauge symmetry restoration in some {\it finite} space-time region as a
result of the presence of the intense gauge pulses.  In some sense, it 
qualitatively resembles the situation realized in the very early Universe 
with the Big Bang scenario where the restoration is caused by the high 
temperature. 
The approach used in this paper might be applied
to the study of chiral symmetry restoration. As is well-known, there exists
an isomorphism between the Higgs sector of the standard model and the linear
$\sigma$-model. This isomorphism is realized through the following 
identifications: $W_L\leftrightarrow\pi$, $H\leftrightarrow\sigma$, $v
\leftrightarrow f_{\pi}$, where $W_L$ is the longitudinal gauge field and $H$ 
the physical Higgs field. $f_{\pi}$ is the $\pi$-decay constant defining the
scale of the chiral dynamics. As a consequence of this isomorphism,
one might expect that the spontaneously broken chiral symmetry can also be
restored in the presence of a strong external pion field and such a physical
setting
might be found in heavy-ion collisions or in the interior of nuclei and
neutron stars.    

We have seen in Section V that the restoration of the 
gauge symmetry in $W$-wave packet collisions occurs at amplitudes for which
strongly inelastic behavior has also been observed.  To achieve a sufficient 
value of 
the amplitude is not easy in practice.  The problem, presumably, is 
connected with the general problem of the classical (quasi-classical) 
treatment of the amplitude for the transition from a few-particle to a 
many-particle final state (see, e.g. \cite{20}).  If, as conjectured in
\cite{21} (see also \cite{22,23,24}), there exist classical trajectories 
interpolating between incoming few-particle and outgoing multi-particle
states, our results would lead us to expect that these trajectories will
also exhibit the phenomenon of symmetry restoration in the transition
region. It is also worthwhile
mentioning that the classical approach is more adequate in the context
discussed here, where the 
colliding 
intense, energetic $W$-boson wave packets serve as a classical external 
source exciting the ground state of the Yang-Mills-Higgs system, and 
we are not interested in the specific nature of the final states.

In conclusion, we emphasize that the restoration of the gauge symmetry
discussed above does not depend on the dimensionality of space.  It is 
expected that the symmetry restoration would not persist as long in 
three space dimensions because the wave packets disperse more rapidly after 
the collisions, causing the average squared amplitude $\langle (W^a_i)^2 \rangle$ 
to decay more rapidly.  The same arguments can apply to the problem of 
multiparticle production in wave packet collisions \cite{6,7}.  This issue, 
as well as the problem of the reliability of the classical treatment of the 
(few$\to$ many) particle amplitude makes the study of the gauge wave packet 
collisions in $(3+1)$ dimensions desirable.

\acknowledgements
We thank K. Rajagopal and R. Singleton for fruitful discussions and
comments.  This work was supported, in part, by grant DE-FG02-96ER40945 
from the U. S. Department of Energy, and by the North Carolina 
Supercomputing Center.

\begin{figure}
\def\epsfsize#1#2{.80#1}
\epsfbox{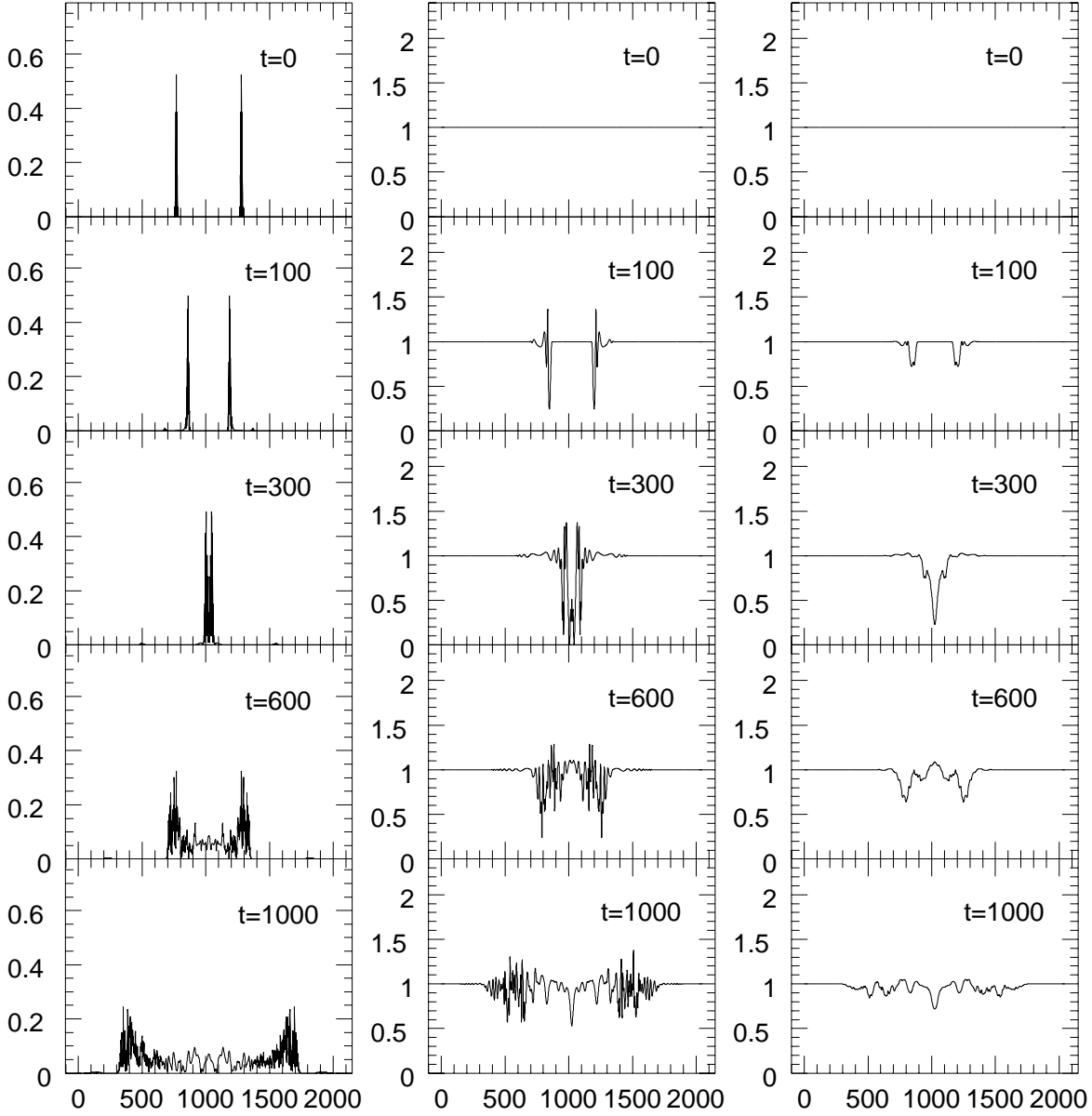}
\bigskip
\caption{Space-time development of symmetry restoration induced by two
colliding gauge wave packets with orthogonal isospins in the presence
of the Higgs vacuum condensate.  First column shows the scaled gauge
field $\vert{\bf A}\vert/v$ as a function of space coordinate $z$ at 
five chosen times.  Second column demonstrates the corresponding 
space-time evolution of the scaled Higgs field $\vert\Phi\vert/v$.  
Third column shows the scaled Higgs field after smoothing over 50
lattice sites.  This simulation as well as the following (Fig. 2) was
done on a lattice of sites $N=2048$ and lattice spacing $a=1$.  The
parameters were $\bar k=\pi/4$, $\Delta k=\pi/16$, $M_H = M_W =0.15$,
$g=0.65$, and $\sigma=1$.}
\end{figure}

\begin{figure}
\def\epsfsize#1#2{.80#1}
\epsfbox{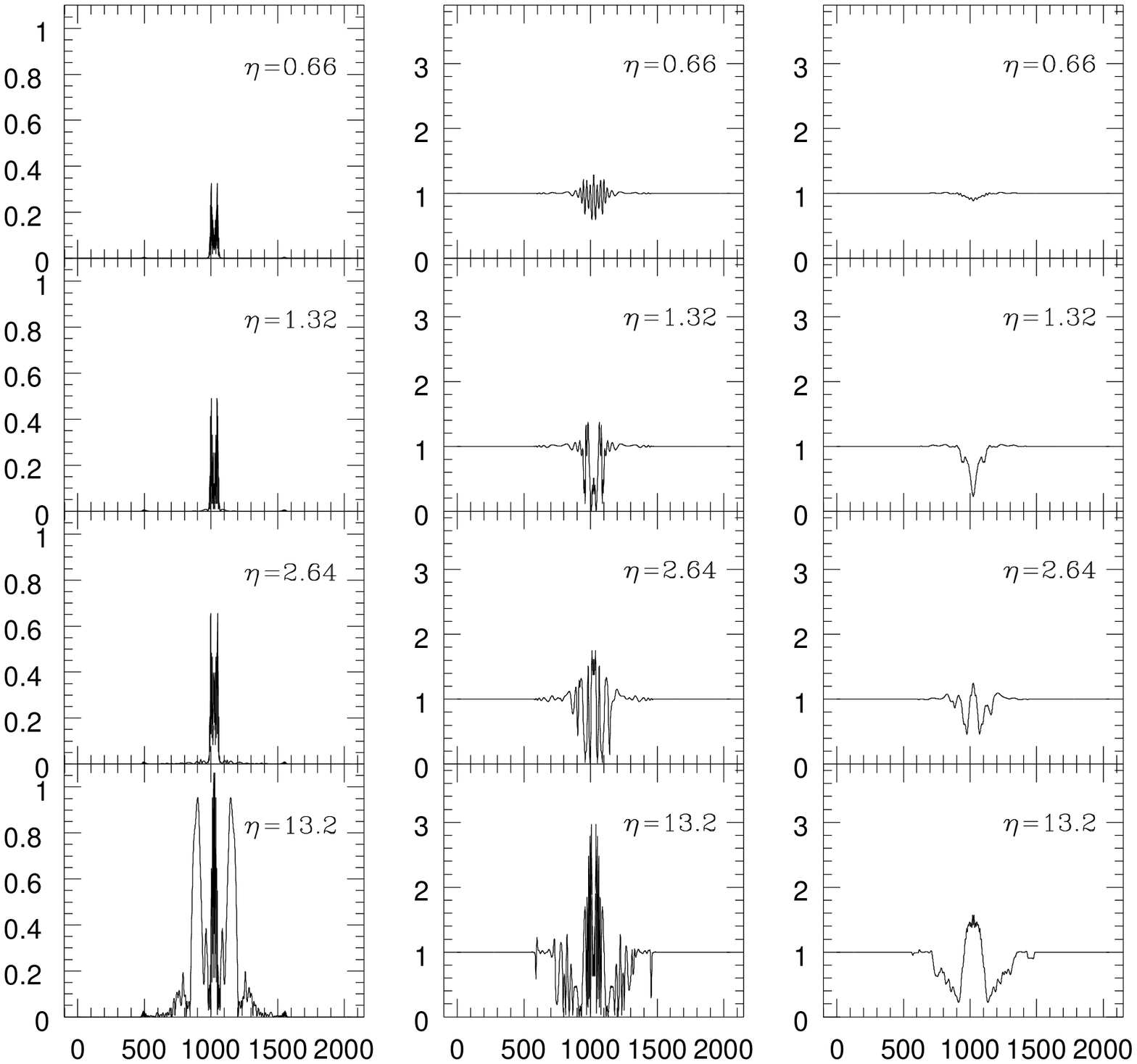}
\caption{$\eta$-dependence of the symmetry restoration.  We demonstrate 
at a chosen time $t=300$ the space pictures of the colliding gauge wave 
packets and the Higgs field for four different values of $\eta$.
First, second, and third columns shows the scaled gauge field 
$\vert{\bf A}\vert/v$, the scaled Higgs field $\vert\Phi\vert/v$, and 
the smoothed
Higgs field $\vert\Phi\vert/v$, respectively, each as a function of the
space coordinate $z$.  All parameters were the same as in Fig. 1 except
$\sigma$, which was to varied to arrive at different values of $\eta$.}
\end{figure}


\begin{references}

\bibitem{1} D. A. Kirzhnits, {\sl JETP Lett. {\bf 15}}, 529 (1972).

\bibitem{2} D. A. Kirzhnits and A. D. Linde, {\sl Phys. Lett. {\bf 42B}},
471 (1972).

\bibitem{3} D. A. Kirzhnits and A. D. Linde, {\sl Ann. Phys. (NY) {\bf 101}},
195 (1976).

\bibitem{4} A. D. Linde, {\sl Rep. Prog. Phys. {\bf 42}}, 389 (1979).

\bibitem{5} I. V. Krive, V. M. Pyzh, and E. M. Chudnovskii, {\sl Sov.
J. Nucl. Phys. {\bf 23}}, 258 (1976).

\bibitem{6} C. R. Hu, S. G. Matinyan, B. M\"uller, A. Trayanov, T. M. Gould,
S. D. Hsu, and E. Poppitz, {\sl Phys. Rev. {\bf D52}}, 2402 (1995).

\bibitem{7} C. R. Hu, S. G. Matinyan, B. M\"uller and D. Sweet,
DUKE-TH-95-97, to appear in {\sl Phys. Rev. {\bf D}}.

\bibitem{8} M. S. Chanowitz and M. K. Gaillard, {\sl Phys. Lett. {\bf B142}},
85 (1984); S. Dawson, {\sl Nucl. Phys. {\bf B249}}, 42 (1985).

\bibitem{9} J. M. Cornwall, D. N. Levin, and G. Tiktopoulos, {\sl Phys.
Rev. {\bf D10}}, 1145 (1974); B. W. Lee, C. Quigg, and M. B. Thacker, 
{\sl Phys. Rev. {\bf D16}}, 1519 (1977); M. S. Chanowitz and M. K.
Gaillard, {\sl Nucl. Phys. {\bf B261}}, 379 (1985).

\bibitem{10} L. S. Brown, {\sl Phys. Rev. {\bf D46}}, R4125 (1992).

\bibitem{11} S. Weinberg, {\sl Phys. Rev. {\bf D9}}, 3357 (1974).

\bibitem{12} L. Dolan and R. Jackiw, {\sl Phys. Rev. {\bf D9}}, 3320 (1974).

\bibitem{13} D. A. Kirzhnits and A. D. Linde, {\sl Sov. Phys. JETP 
{\bf 40}}, 628 (1975).

\bibitem{14} F. Abe et al. CDF Collaboration, {\sl Phys. Rev. Lett. {\bf 74}},
2626 (1995); {\sl Phys. Rev. {\bf D52}}, R2605 (1995); S. Abachi et al. 
D0 Collaboration, {\sl Phys. Rev. Lett. {\bf 74}}, 2632 (1955).

\bibitem{15} S. Coleman and E. Weinberg, {\sl Phys. Rev. {\bf D7}}, 1888
(1973).

\bibitem{16} J.-F. Grivaz, Proceedings of the Intern. EPS Conference on
High-Energy Physics, Bruxelles, 1995.

\bibitem{17} M. Sher, {\sl Physics Reports {\bf 179}}, 273 (1989).

\bibitem{18} J. Kogut and L. Susskind, {\sl Phys. Rev. {\bf D11}}, 395
(1975).

\bibitem{19} T. S. Bir\'o, C. Gong, B. M\"uller, and A. Trayanov, {\sl
Int. J. Mod. Phys. {\bf C5}}, 113 (1994).

\bibitem{20} M. B. Voloshin, in {\sl Proceedings of the 27th International
Conference on High Energy Physics}, Glasgow, Scotland, 1994; v. 1, p. 121.

\bibitem{21} V. Rubakov, D. Son, and P. Tinyakov, 
{\sl Phys. Lett. {\bf B287}}, 342 (1992).

\bibitem{22} T. M. Gould, S. D. H. Hsu, and E. R. Poppitz, 
{\sl Nucl. Phys. {\bf B437}}, 83 (1995).

\bibitem{23} C. Rebbi and R. Singleton, preprint $<$hep-ph/9601260$>$,
Boston University (1996), to appear in {\sl Phys. Rev. {\bf D}}. 

\bibitem{24} E. Farhi, J. Goldstone, A. Lue, and K. Rajagopal, 
{\sl MIT preprint CTP - 2483}, 1995.

\end{references}
\end{document}